\documentclass[rmp,nofootinbib,amsmath,amssymb,twocolumn]{revtex4}
\usepackage{graphicx}
\begin{document}
\title{\bf Origin and Effects of Anomalous Dynamics on Unbiased Polymer
Translocation}
\author{Debabrata Panja} \affiliation{Institute for Theoretical
Physics, Universiteit van Amsterdam, Valckenierstraat 65, 1018 XE
Amsterdam,\\ The Netherlands} \author{Gerard T. Barkema}
\affiliation{Institute for Theoretical Physics, Universiteit Utrecht,
Minnaertgebouw, Leuvenlaan 4, Postbus 80.195,\\ 3508 TD Utrecht, The
Netherlands} \author{Robin C. Ball} \affiliation{Department of
Physics, University of Warwick, Coventry CV4 7AL, UK}
%
%
%
\begin{abstract}
In this paper, we investigate the microscopic dynamics of a polymer of
length $N$ translocating through a narrow pore. Characterization of
its purportedly anomalous dynamics has so far remained incomplete. We show
that the polymer dynamics is anomalous until the Rouse time
$\tau_{R}\sim N^{1+2\nu}$, with a mean square displacement through the
pore consistent with $t^{(1+\nu)/(1+2\nu)}$, with $\nu\approx0.588$
the Flory exponent. This is shown to be directly related to a decay in
time of the excess monomer density near the pore as
$t^{-(1+\nu)/(1+2\nu)}\exp(-t/\tau_{R})$. Beyond the Rouse time
translocation becomes diffusive. In consequence of this, the dwell-time
$\tau_{d}$, the time a translocating polymer typically spends within
the pore, scales as $N^{2+\nu}$, in contrast to previous claims.
\end{abstract}


\maketitle
\section{Introduction\label{sec0}}
Transport of molecules across cell membranes is an essential mechanism
for life processes. These molecules are often long and flexible, and
the pores in the membranes are too narrow to allow them to pass
through as a single unit. In such circumstances, the passage of a
molecule through the pore --- i.e.\ its translocation --- proceeds
through a random process in which polymer segments sequentially move
through the pore.  DNA, RNA and proteins are naturally occurring long
molecules \cite{drei,henry,akimaru,goerlich,schatz}  subject to
translocation in a variety of biological processes. Translocation is
used in gene therapy \cite{szabo,hanss}, in delivery of drug molecules
to their activation sites \cite{tseng}, and as an efficient means of
single molecule sequencing of DNA and RNA
\cite{nakane}. Understandably, the process of translocation has been
an active topic of current research: both because it is an essential
ingredient in many biological processes and for its relevance in
practical applications.

Translocation is a complicated process in living organisms --- its
dynamics may be strongly influenced by various factors, such as the
presence of chaperon molecules, pH value, chemical potential
gradients, and assisting molecular motors. It has been studied
empirically in great variety in the biological literature
\cite{wickner,simon}.  Studies of translocation as a
\emph{biophysical} process are more recent. In these, the polymer is
simplified to a sequentially connected string of $N$
monomers. Quantities of interest are the typical time scale for the
polymer to leave a confining cell or vesicle,  the ``escape time''
\cite{sungpark1}, and the typical time scale the polymer spends in the
pore or ``dwell time'', \cite{sungpark2} as a function of chain length
$N$ and other parameters like membrane thickness, membrane adsorption,
electrochemical potential gradient, etc. \cite{lub}.

\begin{figure}[t]
\begin{center}
\includegraphics[width=0.8\linewidth]{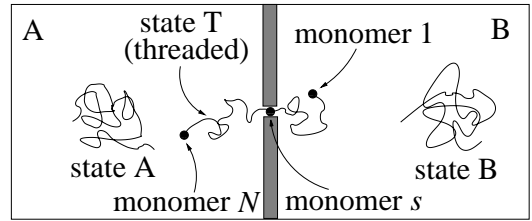}
\end{center}
\caption{Our system to study translocation in this paper. It consists
of two cells A and B that are connected by a pore of diameter unity in
a membrane of thickness unity. Both cells have the same volume $V$
(large compared to the polymer's typical size). The polymer repeatedly
moves back and forth from one cell to the other through the pore. At
any time, exactly one monomer can be within the pore. The Kuhn length
of the polymer and the lattice spacing are also set to unity. Polymers
can be in three different states (i) state A: all monomers are in cell
A; (ii) state T (threaded): some monomers are in cell A and the rest
in cell B; (iii) state B: all monomers are in cell B. The dwell time
$\tau_d$ is defined as the pore-blockade time in experiments, i.e., as how long
the polymer spends in state T during a translocation
event.\label{fig1}}
\end{figure}
These have been measured directly in numerous experiments
\cite{expts}. Experimentally, the most studied quantity is the dwell
time $\tau_{d}$, i.e., the pore blockade time for a translocation
event. For theoretical descriptions of $\tau_{d}$, during the last
decade a number of mean-field type theories
\cite{sungpark1,sungpark2,lub} have been proposed, in which
translocation is described by a Fokker-Planck equation for
first-passage over an entropic barrier in terms of a single ``reaction
coordinate'' $s$. Here $s$ is the number of the monomer threaded at
the pore ($s=1,\ldots,N$). These theories apply under the
assumption that translocation is slower than the equilibration
time-scale of the entire polymer, which is likely for high pore
friction. In Ref.  \cite{kantor}, this assumption was questioned, and
the authors found that for a self-avoiding polymer performing Rouse
dynamics, $\tau_{d}\ge\tau_{R}$, the Rouse time. Using simulation data
in 2D, they suggested that the inequality may actually be an equality,
i.e., $\tau_{d}\sim\tau_{R}\sim N^{1+2\nu}\simeq N^{2.18}$. This
suggestion was numerically confirmed in 2D in Ref. \cite{luo}.
However, in a publication due to two of us, $\tau_{d}$ in 3D was
numerically found to scale as $\sim N^{2.40\pm0.05}$ \cite{wolt}.
Additionally, in a recent publication \cite{dubbeldam} $\tau_{d}$ was
numerically found to scale as $N^{2.52\pm0.04}$ in three dimensions [a
discussion on the theory of Ref. \cite{dubbeldam} appears at the end
of Sec. \ref{sec3}].

Amid all the above results on $\tau_{d}$ mutually differing by
$\sim~O(N^{0.2})$, the only consensus that survives is that
$\tau_{d}\ge\tau_{R}$ \cite{kantor,wolt}.  Simulation results alone
cannot determine the scaling of $\tau_{d}$: different groups use
different polymer models with widely different criteria for
convergence for scaling results, and as a consequence, settling
differences of $\sim~O(N^{0.2})$ in $O(\tau_{R})$, is extremely
delicate.

An alternative approach that can potentially settle the issue of
$\tau_{d}$ scaling with $N$ is to analyze the dynamics of
translocation at a microscopic level. Indeed, the lower limit
$\tau_{R}$ for $\tau_{d}$ implies that the dynamics of translocation
is anomalous \cite{kantor}.  We know of only two published studies on
the anomalous dynamics of translocation, both using a fractional
Fokker-Planck equation (FFPE) \cite{klafter,dubbeldam}. However,
whether the assumptions underlying a FFPE apply for polymer
translocation are not clear. Additionally, none of the studies used
FFPE for the purpose of determining the scaling of $\tau_{d}$. In view
of the above, such a potential clearly has not been thoroughly
exploited.

The purpose of this paper is to report the characteristics of the
anomalous dynamics of translocation, \textit{derived from the
microscopic dynamics of the polymer}, and the scaling of $\tau_{d}$
obtained therefrom. Translocation proceeds via the exchange of
monomers through the pore: imagine a situation when a monomer from the
left of the membrane translocates to the right. This process increases
the monomer density in the right neighbourhood of the pore, and
simultaneously reduces the monomer density in the left neighbourhood
of the pore.  The local enhancement in the monomer density on the
right of the pore \textit{takes a finite time to dissipate away from
the membrane along the backbone of the polymer\/} (similarly for
replenishing monomer density on the left neighbourhood of the
pore). The imbalance in the monomer densities between the two local
neighbourhoods of the pore during this time implies that there is an
enhanced chance of the translocated monomer to return to the left of
the membrane, thereby giving rise to \textit{memory effects\/}, and
consequently, rendering the translocation dynamics subdiffusive. More
quantitatively, the excess monomer density (or the lack of it) in the
vicinity of the pore manifests itself in reduced (or increased) chain
tension around the pore, creating an imbalance of chain tension across
the pore (we note here that the chain tension at the pore acts as
monomeric chemical potential, and from now on we use both terms
interchangeably). We use well-known laws of polymer physics to show
that in time the imbalance in the chain tension across the pore
relaxes as $t^{-(1+\nu)/(1+2\nu)}\exp(-t/\tau_{R})$
\cite{strict}. This results in translocation dynamics being
subdiffusive for $t<\tau_{R}$, with the mean-square displacement
$\langle\Delta s^{2}(t)\rangle$ of the reaction co-ordinate $s(t)$
increasing as $t^{(1+\nu)/(1+2\nu)}$; and diffusive for
$t>\tau_{R}$. With $\sqrt{\langle\Delta s^{2}(\tau_{d})\rangle}\sim
N$, this leads to $\tau_{d}\sim N^{2+\nu}$.

This paper is divided in four sections. In Sec. \ref{sec1} we detail
the dynamics of our polymer model, and outline its implications on
physical processes including equilibration of phantom and
self-avoiding polymers. In Sec. \ref{sec2} we elaborate on a novel way
of measuring the dwell time that allows us to obtain better statistics
for large values of $N$. In Sec. \ref{sec3} we describe and
characterize the anomalous dynamics of translocation, obtain the
scaling of $\tau_d$ with $N$ and compare our results with that of
Ref. \cite{dubbeldam}. In Sec. \ref{sec4} we end this paper with a
discussion.
\begin{figure}[h]
\begin{center}
\includegraphics[width=0.84\linewidth]{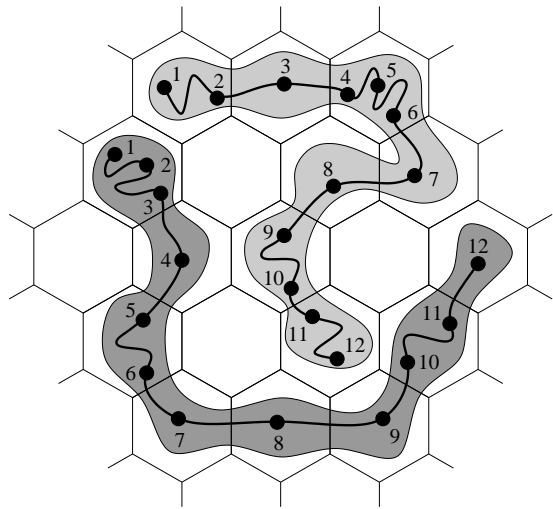}
\end{center}
\caption{Illustration of the two-dimensional version of the lattice
polymer model.  In the upper polymer, interior monomers 2, 4, 6, 9, 10
and 11 can either diffuse along the polymer backbone, or move
sideways; monomer 7 can join either 6 or 8; the end monomers 1 and 12
can move to any vacant nearest-neighbor site.  In the lower polymer,
interior monomers 3, 5, 6, 10 and 11 can either diffuse along the
tube, or move sideways; monomer 1 can move to any vacant
nearest-neighbor site, and monomer 12 can join its neighbor 11. All
other monomers are not mobile.\label{fig2}}
\end{figure}
\begin{figure*}
\begin{center}
\begin{minipage}{0.45\linewidth} 
\includegraphics[width=0.8\linewidth,angle=270]{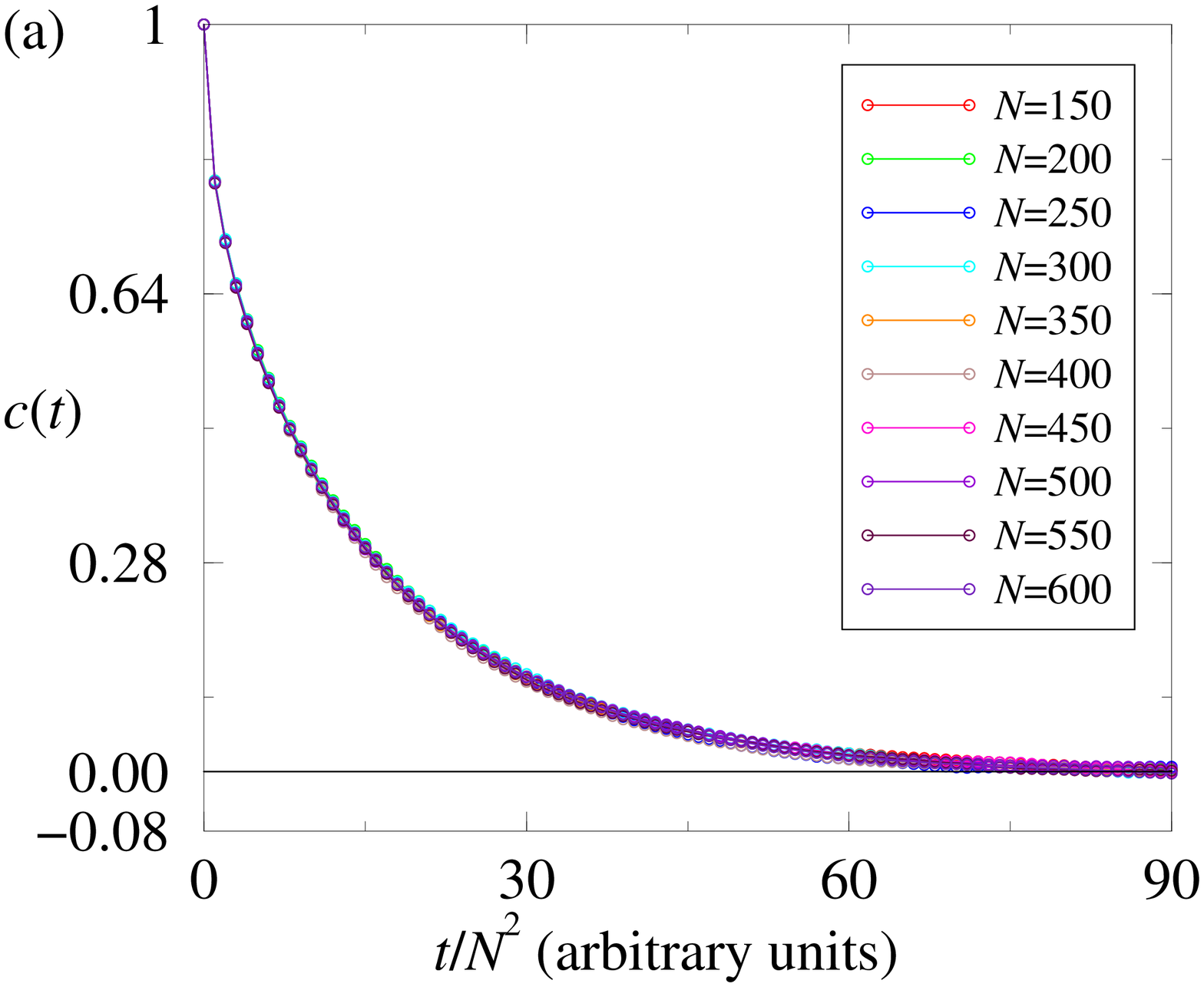}
\end{minipage}
\hspace{10mm}
\begin{minipage}{0.45\linewidth}
\includegraphics[width=0.8\linewidth,angle=270]{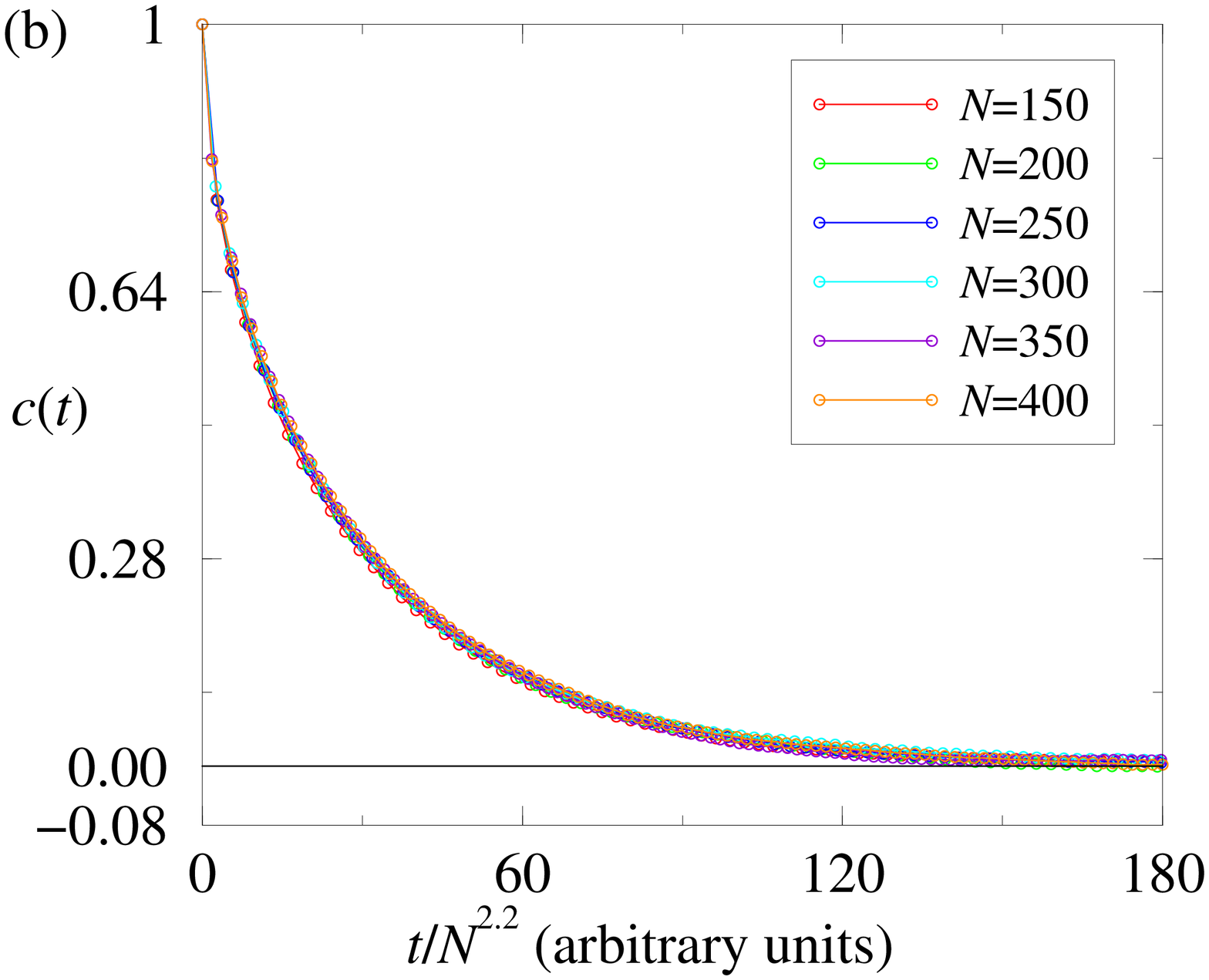}
\end{minipage} 
\end{center}
\caption{Collapse of $c(t)$ for different values of $N$, showing that
the equilibration times for phantom and self-avoiding polymers scale
as $N^2$ and $N^{1+2\nu}$ respectively. Here $N$ is the polymer
length. (a) Data for phantom polymers, simulations were run with the
same definition of time for all values of $N$; to achieve the data
collapse, the times were then scaled by a factor $N^2$. (b) Data for
self-avoiding polymers, simulations were run with the same definition
of time for all values of $N$; to achieve the data collapse, the times
were scaled by a factor $N^{1+2\nu}$ ($N^{2.2}$ to be precise). Note
that the units of time are arbitrary and clearly not important for the
scaling of polymer equilibration times.}
\label{fig3}
\end{figure*}

\section{Our polymer model\label{sec1}}

Over the last years, we have developed a highly efficient simulation
approach of polymer dynamics. This approach is made possible via a
lattice polymer model, based on Rubinstein's repton model
\cite{rubinstein} for a single reptating polymer, with the addition of
sideways moves (Rouse dynamics) and entanglement. A detailed
description of this lattice polymer model, its computationally
efficient implementation and a study of some of its properties and
applications can be found in Refs. \cite{heukelum03,LatMCmodel}.

In this model, polymers consist of a sequential chain of monomers,
living on a FCC lattice. Monomers adjacent in the string are located
either in the same, or in neighboring lattice sites. The polymers are
self-avoiding: multiple occupation of lattice sites is not allowed,
except for a set of adjacent monomers. The polymers move through a
sequence of random single-monomer hops to neighboring lattice
sites. These hops can be along the contour of the polymer, thus
explicitly providing reptation dynamics. They can also change the
contour ``sideways'', providing Rouse dynamics. Time in this polymer
model is measured in terms of the number of attempted reptation
moves. A two-dimensional version of our three-dimensional model is
illustrated in fig. \ref{fig2}.

\subsection{Influence of the accelerated reptation moves on polymer
dynamics\label{sec1.1}}  From our experience with the model we already
know that the dynamical properties are rather insensitive to the ratio
of the rates for Rouse vs. reptation moves (i.e., moves that alter the
polymer contour vs. moves that only redistribute the stored length
along the backbone). Since the computational efficiency of the latter
kind of moves is at least an order of magnitude higher, we exploit
this relative insensitivity by attempting reptation moves $q$ times
more often than Rouse moves; typical values are $q=1$, $5$ or 10, which
correspond to a comparable amount of computational effort invested in
both kinds of moves. Certainly, the interplay between the two kinds of
moves is rather intricate \cite{wolt2}.  Recent work by Drzewinski and
van Leeuwen on a related lattice polymer model \cite{drz} provides
evidence that the dynamics is governed by $Nq^{-1/2}$, supporting our
experience that, provided the polymers are sizable, one can boost one
mechanism over the other quite a bit (even up to $q \sim N^2$) before
the polymer dynamics changes qualitatively.
\begin{figure*}
\begin{center}
\includegraphics[width=0.75\linewidth]{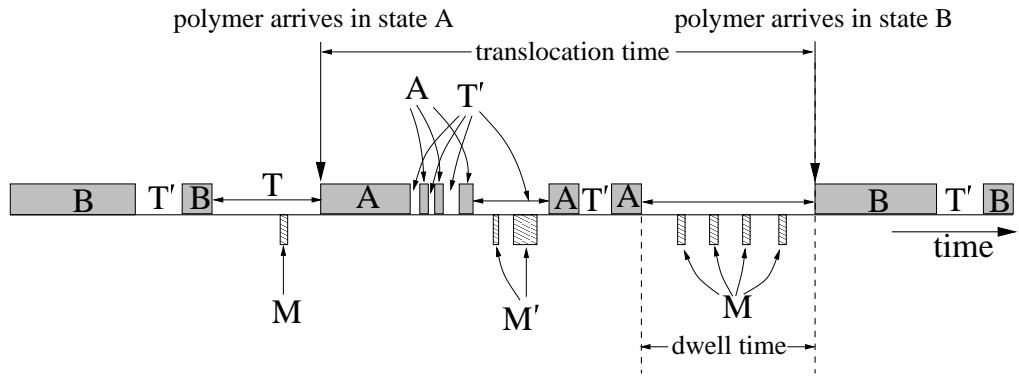}
\caption{A typical translocation process of the polymer as for our
system the polymers move repeatedly back and forth between cells A and
B. See Fig. \ref{fig1} and text for the definitions of the states A,
B, T, M, $\mbox{T}'$ and $\mbox{M}'$.}
\label{fig4}
\end{center}
\end{figure*}
In order to further check the trustworthiness of this model, we use
it to study the equilibration properties of polymers with one end
tethered to a fixed infinite wall (this problem relates rather
directly to that of a translocating polymer: for a given monomer
threaded into the pore, the two segments of the polymer on two sides
of the membrane behave simply as two independent polymer chains; see
Fig. \ref{fig1}). This particular problem, wherein polymer chains (of
length $N$) undergo pure Rouse dynamics (i.e., no additional reptation
moves) is a well-studied one: the equilibration time is known
to scale as $N^{1+2\nu}$ for self-avoiding polymers and as $N^2$ for
phantom polymers. To reproduce these results with our model we denote
the vector distance of the free end of the polymer w.r.t. the
tethered end at time $t$ by $\mathbf{e}(t)$, and define the
correlation coefficient for the end-to-end vector as
\begin{eqnarray}
c(t) =
\frac{\langle\mathbf{e}(t)\cdot\mathbf{e}(0)\rangle-\langle\mathbf{e}(t)\rangle\cdot\langle\mathbf{e}(0)\rangle}{\sqrt{\langle\mathbf{e}^2(t)-\langle\mathbf{e}(t)\rangle^2\rangle\langle\mathbf{e}^2(0)-\langle\mathbf{e}(0)\rangle^2\rangle}}\,.
\label{eq1}
\end{eqnarray}
The angular brackets in Eq. (\ref{eq1}) denote averaging in
equilibrium. The $\bar{c}(t)$ quantities appearing in Fig. \ref{fig3}
have been obtained by the following procedure: we first obtain the
time correlation coefficients $c(t)$ for $32$ independent polymers, and
$\bar{c}(t)$ is a further arithmatic mean of the corresponding $32$
different time correlation coefficients. For different values of $N$
we measure $\bar{c}(t)$ for both self-avoiding and phantom
polymers. When we scale the units of time by factors of $N^{2}$ and
$N^{2.2}$ respectively for phantom and self-avoiding polymers, the
$\bar{c}(t)$ vs. $t$ curves collapse on top of each other. This is
shown in Fig. \ref{fig3}. Note here that $1+2\nu=2.175$, which is
sufficiently close to $2.2$, and in simulations of self-avoiding
polymers [Fig. \ref{fig2}(b)] we cannot differentiate between $1+2\nu$
and $2.2$.

\section{Translocation, dwell and unthreading times\label{sec2}}

Our translocation simulations are carried out only for self-avoiding
polymers. For long polymers, full translocation simulations, i.e.,
having started in one of the cells (Fig. \ref{fig1}), finding the pore
and distorting their shapes to enter the pore to finally translocate
to the other side of the membrane are usually very slow. An uncoiled
state, which the polymer has to get into in order to pass through the
pore is entropically unfavourable, and therefore, obtaining good
translocation statistics for translocation events is a time-consuming
process. To overcome this difficulty, in Ref. \cite{wolt}, we
introduced three different time scales associated with translocation
events: translocation time, dwell time and unthreading time. For the
rest of this section, we refer the reader to Fig. \ref{fig1}.

\subsection{Translocation and dwell times\label{sec2.1}}

In states A and B (Fig. \ref{fig1}), the entire polymer is located in
cell A, resp.\ B. States M and ${\text M}'$ are defined as the states
in which the middle monomer is located exactly halfway between both
cells. Finally, states T and ${\text T}'$ are the complementary to the
previous states: the polymer is threaded, but the middle monomer is
not in the middle of the pore. The finer distinction between states M
and T, resp.\ ${\text M}'$ and ${\text T}'$ is that in the first case,
the polymer is on its way from state A to B or vice versa, while in
the second case it originates in state A or B and returns to the same
state. The translocation process in our simulations can then be
characterized by the sequence of these states in time
(Fig. \ref{fig4}). In this formulation, the dwell time $\tau_d$ is the
time that the polymer spends in states T, while the translocation time
$\tau_t$ is the time starting at the first instant the polymer reaches
state A after leaving state B, until it reaches state B again. As
found in Ref. \cite{wolt}, $\tau_d$ and $\tau_t$ are related to each
other by the relation
\begin{eqnarray}
\tau_t\,=\,V\,\tau_d\,N^{1+\gamma-2\gamma_1}\,,
\label{eq0}
\end{eqnarray}
where $\gamma=1.1601$, $\gamma_1=0.68$ and $V$ is the volume of 
cell A or B (see Fig. \ref{fig1}).

\subsection{Unthreading time and its relation to dwell
  time\label{sec2.2}}

The unthreading time $\tau_u$ is the average time that either state A
or B is reached from state M (not excluding possible recurrences of
state M). Notice that in the absence of a driving field, on average,
the times to unthread to state A equals the the times to unthread to
state B, due to symmetry. The advantage of introducing unthreading
time is that when one measures the unthreading times, the polymer is
at the state of lowest entropy at the start of the simulation and
therefore simulations are fast and one can obtain good statistics on
unthreading times fairly quickly. Additionally, the dwell and
unthreading times are related to each other, as outlined below, and
using this relation one is able to reach large values of $N$ for
obtaining the scaling of the dwell time.

The main point to note that the dwell time can be decomposed into
three parts as
\begin{eqnarray}
\tau_d=\tau_{\mbox{\scriptsize
A}\to\mbox{\scriptsize M}}\,+\,\tau_{\mbox{\scriptsize
M}\looparrowright\mbox{\scriptsize
M}}\,+\,\tau_{\mbox{\scriptsize M}\to\mbox{\scriptsize
B}}\,,
\label{eq2}
\end{eqnarray}
whereas mean unthreading time can be decomposed into two parts as
\begin{eqnarray}
\tau_u=\tau_{\mbox{\scriptsize
M}\looparrowright\mbox{\scriptsize
M}}\,+\,\tau_{\mbox{\scriptsize M}\to\mbox{\scriptsize
B}}\,.
\label{eq3}
\end{eqnarray}
Here $\tau_{\mbox{\scriptsize A}\to\mbox{\scriptsize
M}}$, $\tau_{\mbox{\scriptsize
M}\looparrowright\mbox{\scriptsize M}}$ and
$\tau_{\mbox{\scriptsize M}\to\mbox{\scriptsize B}}$
respectively are the mean first passage time to reach state M from
state A, mean time between the first occurance at state M and the last
occurance of state M with possible reoccurances of state M, and the
mean first passage time to reach state B from state M without any
reoccurance of state M. Since on average $\tau_{\mbox{\scriptsize
A}\to\mbox{\scriptsize M}}=\tau_{\mbox{\scriptsize
M}\to\mbox{\scriptsize B}}$ due to symmetry, and all quantities
on the r.h.s. of Eqs. (\ref{eq2}) and (\ref{eq3}) are strictly
positive, we arrive at the inequality
\begin{eqnarray}
\tau_u<\tau_d<2\tau_u\,.
\label{eq4}
\end{eqnarray}
Since Eq. (\ref{eq4}) is independent of polymer lengths, on average,
the dwell time scales with $N$ in the same way as the unthreading time
\cite{mistake}.

\section{Characterization of the anomalous dynamics of translocation
  and its relation to $\tau_d$\label{sec3}} 

The reaction coordinate [the monomer number $s(t')$, which is occupying
the pore at time $t'$] a convenient choice for the description of the
microscopic movements of the translocating polymer, since the
important time-scales for translocation can be obtained from the time
evolution of the reaction coordinate $s(t')$ as shown below. To delve
deeper into its temporal behavior, we determine
$P_{N,r}(s_1,s_2,t)$, the probability distribution that at time
$t$ a polymer of length $N$ is in a configuration for which monomer
$s_1$ is threaded into the pore, and the polymer evolves at time
$t'+t$ into a configuration in which monomer $s_2$ is threaded
into the pore. The subscript $r$ denotes our parametrization for the
polymer movement, as it determines the frequency of attempted
reptation moves and the sideways (or Rouse) moves of the polymer. See
Sec. \ref{sec1.1} for details.

To maintain consistency, all simulation results reported here in this
section are for $q=10$, so we drop $q$ from all notations from here
on. This value for $q$ is used in view of our experience with the
polymer code: $q=10$ yields the fastest runtime of our code; this was
the same value used in our earlier work \cite{wolt}. As discussed in
Sec. \ref{sec1.1}, this value of $q$ does not change any physics.

\begin{figure}[h]
\begin{center}
\includegraphics[width=0.8\linewidth,angle=270]{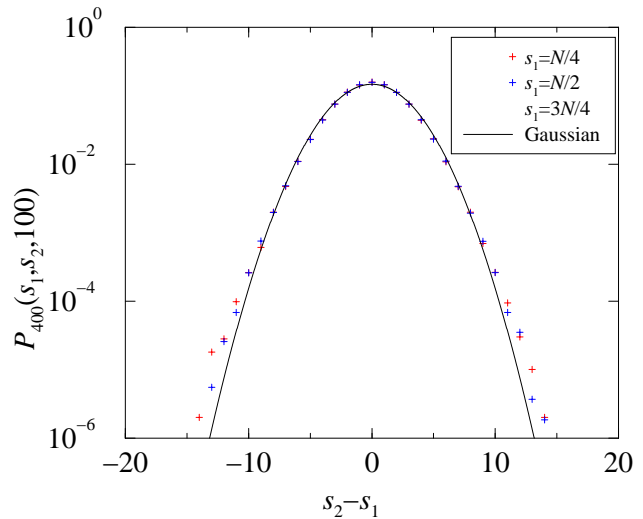}
\end{center}
\caption{$P_{N}(s_1,s_2,t)$  for $N=400$, at
$t=100$. Note the data collapse when plotted as a function of
$(s_2-s_1)$. The distribution differs slightly from
Gaussian. \label{fig5}}
\end{figure}

First, we investigate the shape of these probability distributions for
various values of $s_1$, $s_2$ and $t$, for  different sets of $N$ and
$r$ values. We find that as long as the $t$ values are such that
neither $s_1$ nor $s_2$ are too close to the end of the polymer,
$P_{N}(s_1,s_2,t)$ depends only on $(s_2-s_1)$, but the centre of the
distribution is slightly shifted w.r.t. the starting position $s_1$ by
some distance that depends on $s_1$, $N$ and $q$. This is illustrated
in Fig. \ref{fig5}, where (on top of each other) we plot
$P_{N}(s_1,s_2,t)$ for $s_1=N/4, N/2$ and $3N/4$ at $t=100$ time
units, for $N=400$, as well as the Gaussian distribution. Notice in
fig. \ref{fig5} that the distribution $P_{N}(s_1,s_2,t)$ differs
slightly from Gaussian (the parameter for the Gaussian is calculated
by least-square optimization). We find that this difference decreases
with increasing values of $t$ (not shown in this paper).

We now define the mean  $\langle[s_2-s_1]\rangle(s_1,t)$ and the
variance $\langle\Delta s^2(s_1,t)\rangle$ of the distribution
$P_{N}(s_1,s_2,t)$ as
\begin{eqnarray}
\langle[s_2-s_1]\rangle(s_1,t)\,=\,\int
ds_2\,P_{N}(s_1,s_2,t)\,(s_2-s_1); \nonumber \\
&&\hspace{-7.85cm}\langle\Delta s^2(s_1,t)\rangle\,=\,\int ds_2\,P_{N}(s_1,s_2,t)\times\nonumber\\
&&\hspace{-5.8cm}\{(s_2-s_1)^2-\,[\langle s_2-s_1\rangle(s_1,t)]^2\}\,,
\label{eq5}
\end{eqnarray}
where the quantities within parenthesis on the l.h.s. of
Eq. (\ref{eq5}) denote the functional dependencies of the mean and the
standard deviation of $(s_2-s_1)$. We also note here that we have
checked the skewness of $P_{N}(s_1,s_2,t)$, which we found to
be zero within our numerical abilities, indicating that
$P_{N}(s_1,s_2,t)$ is symmetric in $(s_2-s_1)$.

In principle, both the mean and the standard deviation of $(s_2-s_1)$
can be used to obtain the scaling of $\tau_d$ with $N$, but the
advantage of using $\langle\Delta s^2(s_1,t)\rangle$ for this purpose is
that, as shown in Fig. \ref{fig5}, it is independent of $s_1$, so from
now on, we drop $s_1$ from its argument. Since
unthreading process starts at $s_1=N/2$, the scaling of
$\tau_d$ with $N$ is easily obtained by using the
relation
\begin{eqnarray}
\langle\Delta s^2(\tau_d)\rangle\sim N^2\,,
\label{eq6}
\end{eqnarray}
in combination with the fact that the scalings of $\tau_d$ and
$\tau_u$ with $N$ behave in the same way [inequality
(\ref{eq4})]. Note here that Eq. (\ref{eq6}) uses the fact that for an
unthreading process the polymer only has to travel a length $N/2$
along its contour in order to leave the pore.

\subsection{The origin of anomalous dynamics and the relaxation of
excess monomer density near the pore during translocation \label{sec3.1}}

The key step in quantitatively formulating the anomalous dynamics of
translocation is the following observation: a translocating polymer
comprises of \textit{two polymer segments tethered at opposite ends of
the pore\/} that are able to exchange monomers between them through
the pore; so \textit{each acts as a reservoir of monomers for the
other.} The velocity of  translocation $v(t)=\dot{s}(t)$, representing
monomer current, responds to $\phi(t)$, the imbalance in the monomeric
chemical potential across the pore acting as
``voltage''. Simultaneously, $\phi(t)$ also adjusts in response to
$v(t)$. In the presence of memory effects, they are related
to each other by $\phi(t)=\int_{0}^{t}dt'\mu(t-t')v(t')$ via the
memory kernel $\mu(t)$, which can be thought of as the
(time-dependent) `impedance' of the system. Supposing a zero-current
equilibrium condition at time $0$, this relation can be
inverted to obtain $v(t)=\int_{0}^{t}dt'a(t-t')\phi(t')$, where $a(t)$
can be thought of as the `admittance'. In the Laplace transform
language, $\tilde{\mu}(k)=\tilde{a}^{-1}(k)$, where $k$ is the Laplace
variable representing inverse time. Via the fluctuation-dissipation
theorem, they are related to the respective autocorrelation
functions as $\mu(t-t')=\langle\phi(t)\phi(t')\rangle_{v=0}$ and
$a(t-t')=\langle v(t)v(t')\rangle_{\phi=0}$.

\begin{figure}[!h]
\includegraphics[width=0.8\linewidth,angle=270]{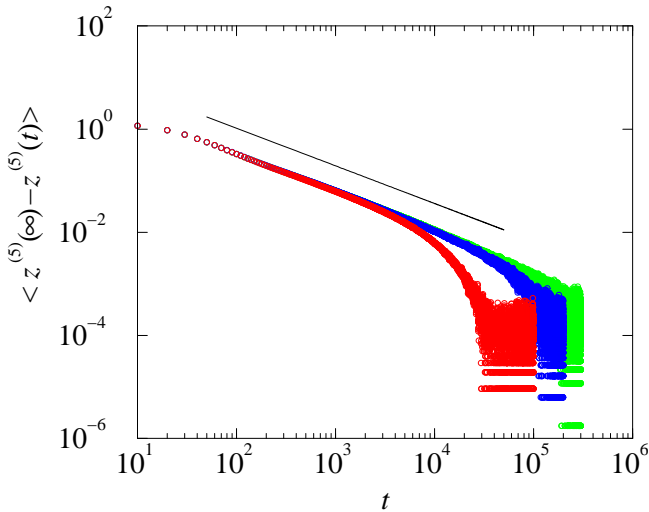} 
\caption{(colour online) Simulation results for the average chain
tension component perpendicular to the membrane proxied by $\langle
z^{(5)}(\infty)-z^{(5)}(t)\rangle$ following monomer injection at the
pore corresponding to $v(t)=p\delta(t)$, with $p=10$. See text for
details. Red circles: $N/2=50$, blue circles: $N/2=100$, green
circles: $N/2=150$, solid black line: $t^{-(1+\nu)/(1+2\nu)}$ with
$\nu=0.588$ for self-avoiding polymers. The steeper drop at large
times correspond to the exponential decay $\exp(-t/\tau_{R})$ (we have
checked this, but have not shown in this letter). \label{fig6}}
\end{figure}
The behaviour of $\mu(t)$ may be obtained by considering the polymer
segment on one side of the membrane only, say the right, with a sudden
introduction of $p$ extra monomers at the pore, corresponding to
impulse current $v(t)=p\delta(t)$. We then ask for the time-evolution
of the mean response $\langle\delta\Phi^{(r)}(t)\rangle$, where
$\delta\Phi^{(r)}(t)$ is the shift in chemical potential for the
right segment of the polymer at the pore. This means that for the
translocation problem (with both right and left segments), we would
have $\phi(t)=\delta\Phi^{(r)}(t)-\delta\Phi^{(l)}(t)$, where
$\delta\Phi^{(l)}(t)$ is the shift in chemical potential for the
left segment at the pore due to an opposite input current to it.

We now argue that this mean response, and hence $\mu(t)$, takes the
form $\mu(t)\sim t^{-\alpha}\exp(-t/\tau_{R})$. The terminal
exponential decay $\exp(-t/\tau_{R})$ is expected from the relaxation
dynamics of the entire right segment of the polymer with one end
tethered at the pore [see Fig. \ref{fig3}(b)]. To understand the
physics behind the exponent $\alpha$, we use the well-established
result for the relaxation time $t_{n}$ for $n$ self-avoiding Rouse
monomers scaling as $t_{n}\sim n^{1+2\nu}$. Based on the expression of
$t_{n}$, we anticipate that by time $t$ the extra monomers will be
well equilibrated across the inner part of the chain up to $n_{t}\sim
t^{1/(1+2\nu)}$ monomers from the pore, but not significantly
further. This internally equilibrated section of $n_{t}+p$ monomers
extends only $r(n_{t})\sim n_{t}^{\nu}$, less than its equilibrated
value $\left(n_{t}+p\right)^{\nu}$, because the larger scale
conformation has yet to adjust: the corresponding compressive force
from these $n_{t}+p$ monomers is expected by standard polymer scaling
\cite{degennes} to follow $f/(k_{B}T)\sim\delta
r(n_{t})/r^{2}(n_{t})\sim\nu p/\left[n_{t}r(n_{t})\right]\sim
t^{-(1+\nu)/(1+2\nu)}$.  This force $f$ must be transmitted to the
membrane, through a combination of decreased tension at the pore and
increased incidence of other membrane contacts. The fraction borne by
reducing chain tension at the pore leads us to the inequality
$\alpha\ge(1+\nu)/(1+2\nu)$, which is significantly different from
(but compatible with) the value $\alpha_{1}=2/(1+2\nu)$ required to
obtain $\tau_{d}\sim\tau_{R}$ . It seems unlikely that the adjustment
at the membrane should be disproportionately distributed between the
chain tension at the pore and other membrane contacts, leading to the
expectation that the inequality above is actually an equality.

We have confirmed this picture by measuring the impedance response
through simulations. In Ref. \cite{forced}, two of us have shown that
the centre-of-mass of the first few monomers is an excellent proxy for
chain tension at the pore and we assume here that this further
serves as a proxy for $\delta\Phi$. Based on this idea, we track
$\langle\delta\Phi^{(r)}(t)\rangle$ by measuring the distance of the average
centre-of-mass of the first $5$ monomers from the membrane, $\langle
z^{(5)}(t)\rangle$, in response to the injection of extra monomers
near the pore at time $0$. Specifically we consider the equilibrated
right segment of the polymer, of length $N/2-10$ (with one end
tethered at the pore), adding $p=10$ extra monomers at the tethered
end of the right segment at time $0$, corresponding to $p=10$,
bringing its length up to $N/2$. Using the proxy $\langle
z^{(5)}(t)\rangle$ we then track $\langle\delta\Phi^{(r)}(t)\rangle$. The
clear agreement between the exponent obtained from the simulation
results with the theoretical prediction of $\alpha=(1+\nu)/(1+2\nu)$
can be seen in Fig. \ref{fig6}. We have checked that the sharp
deviation of the data from the power law $t^{-(1+\nu)/(1+2\nu)}$ at
long times is due to the asymptotic exponential decay as
$\exp(-t/\tau_{R})$, although this is not shown in the figure.

Having thus shown that $\mu(t)\sim
t^{-\frac{1+\nu}{1+2\nu}}\exp(-t/\tau_{R})$, we can expect that the
translocation dynamics is anomalous for $t<\tau_{R}$, in the sense
that the mean-square displacement of the monomers through the pore,
$\langle\Delta s^{2}(t)\rangle\sim t^{\beta}$ for some $\beta<1$ and
time $t<\tau_{R}$, whilst beyond the Rouse time it becomes simply
diffusive. The value $\beta=\alpha=\frac{1+\nu}{1+2\nu}$ follows
trivially by expressing $\langle\Delta s^{2}(t)\rangle$ in terms of
(translocative) velocity correlations $\left\langle
v(t)v(t')\right\rangle $, which (by the Fluctuation Dissipation
theorem) are given in terms of the time dependent admittance
$a(t-t')$.  and hence inversely in terms of the corresponding
impedance.

\begin{figure}[!h]
 \begin{centering}
\includegraphics[angle=270,width=0.95\linewidth]{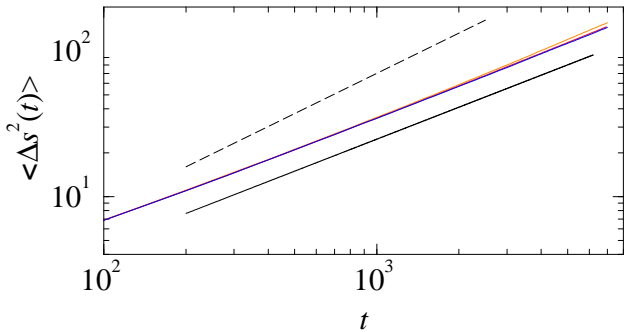} 
\par\end{centering}
\caption{(colour online) Double-logarithmic plot of the mean squared
displacement of the reaction coordinate $\langle\Delta
s^{2}(t)\rangle$ as a function of time $t$, for $N=100$ (orange), 200
(red) and 500 (blue). The thick black line indicates the theoretically
expected slope corresponding to $\langle\Delta s^{2}(t)\rangle\sim
t^{(1+\nu)/(1+2\nu)}$. The dashed black line corresponds to
$\langle\Delta s^{2}(t)\rangle\sim t^{2/(1+2\nu)}$, which would have
been the slope of the $\langle\Delta s^2(t)\rangle$ vs. $t$ curve in
a double-logarithmic plot, if $\tau_d$ were to scale as $\tau_R\sim
N^{1+2\nu}$.
\label{fig7}}
\end{figure}
Indeed, as shown in Fig. \ref{fig7}, a double-logarithmic plot of
$\langle\Delta s^{2}(t)\rangle$ vs. $t$ is consistent with
$\langle\Delta s^{2}(t)\rangle\sim t^{(1+\nu)/(1+2\nu)}$. The
behaviour of $\langle\Delta s^{2}(t)\rangle$ at short times is an
artifact of our model: at short times reptation moves dominate,
leading to a transport mechanism for {}``stored lengths''
\cite{rubinstein} along the polymer's contour in which individual
units of stored length cannot pass each other. As a result, the
dynamics of $s(t)$, governed by the movement of stored length units
across the pore, is equivalent to a process known as {}``single-file
diffusion'' on a line, characterized by the scaling $\langle\Delta
s^{2}(t)\rangle\sim t^{1/2}$ (not shown here). At long times the
polymer tails will relax, leading to $\langle\Delta
s^{2}(t)\rangle\sim t$ for $t>\tau_{R}$. The presence of two
crossovers, the first one from $\langle\Delta s^{2}(t)\rangle\sim
t^{1/2}$ to $\langle\Delta s^{2}(t)\rangle\sim t^{(1+\nu)/(1+2\nu)}$
and the second one from $\langle\Delta s^{2}(t)\rangle\sim
t^{(1+\nu)/(1+2\nu)}$ to $\langle\Delta s^{2}(t)\rangle\sim t$ at
$t\approx\tau_{R}$, complicates the precise numerical verification of
the exponent $(1+\nu)/(1+2\nu)$. However, as shown in Fig. \ref{fig7},
there is an extended regime in time at which the quantity
$t^{-(1+\nu)/(1+2\nu)}\langle\Delta s^{2}(t)\rangle$ is nearly
constant.

The subdiffusive behaviour $\langle\Delta s^{2}(t)\rangle\sim
t^{\frac{1+\nu}{1+2\nu}}$ for $t<\tau_{R}$, combined with the
diffusive behaviour for $t\geq\tau_{R}$ leads to the dwell time
scaling as $\tau_{d}\sim N^{2+\nu}$, based on the criterion that
$\sqrt{\langle\Delta s^{2}(\tau_{d})\rangle}\sim N$.  The dwell time
exponent $2+\nu\simeq2.59$ is in acceptable agreement with the two
numerical results on $\tau_{d}$ in 3D as mentioned in the introduction
of this letter, and in Table I below we present new high-precision
simulation data in support of $\tau_{d}\sim N^{2+\nu}$, in terms of
the median unthreading time. The unthreading time $\tau_u$ is defined
as the time for the polymer to leave the pore with $s(t=0)=N/2$ and
the two polymer segments equilibrated at $t=0$. Both $\tau_u$ and
$\tau_d$ scale the same way, since $\tau_u<\tau_d<2\tau_u$
[see Eq. (\ref{eq4})].
\begin{center}
\begin{tabular}{p{2cm}|p{2cm}|p{2cm}}
\hspace{8mm}$N$ &
\hspace{8mm}$\tau_u$ &
\hspace{4mm}$\tau_u/N^{2+\nu}$ \tabularnewline
\hline
\hline 
\hspace{7mm}100 &
\hspace{6mm}65136 &
\hspace{6mm}0.434 \tabularnewline
\hline 
\hspace{7mm}150 &
\hspace{5mm}183423 &
\hspace{6mm}0.428 \tabularnewline
\hline 
\hspace{7mm}200 &
\hspace{5mm}393245 &
\hspace{6mm}0.436 \tabularnewline
\hline 
\hspace{7mm}250 &
\hspace{5mm}714619 &
\hspace{6mm}0.445 \tabularnewline
\hline 
\hspace{7mm}300 &
\hspace{4mm}1133948&
\hspace{6mm}0.440\tabularnewline
\hline 
\hspace{7mm}400 &
\hspace{4mm}2369379&
\hspace{6mm}0.437\tabularnewline
\hline 
\hspace{7mm}500 &
\hspace{4mm}4160669&
\hspace{6mm}0.431\tabularnewline
\hline
\end{tabular}
\par\end{center}
{\footnotesize Table I: Median unthreading time over 1,024 runs for
each $N$.} \vspace{3mm}

Our results have two main significant implications:
\begin{itemize}
\item[(i)] Even in the limit of small (or negligible) pore friction,
equilibration time scale of a polymer is smaller than its dwell time
scale. Yet, quasi-equilibrium condition cannot be assumed as
the starting point to analyze the dynamics of unbiased translocation,
as has been done in the mean-field theories.

\item[(ii)] Since $\alpha=(1+\nu)/(1+2\nu)<1$, the diffusion in
reaction-coordinate space is anomalous. This means that the dynamics
of the translocating polymer in terms of its reaction co-ordinate {\it
cannot\/} be captured by a Fokker-Planck equation in the limit of
small (or negligible) pore friction.

In view of our results, a Fokker-Planck type equation [such as a
fractional Fokker-Planck equation (FFPE)] to describe the anomalous
dynamics of a translocating polymer would definitely need input from
the physics of polymer translocation. It therefore remains to be
checked that the assumptions underlying a FFPE does not violate the
basic physics of a translocating polymer.
\end{itemize}

\subsection{Comparison of our results with the theory of
  Ref. \cite{dubbeldam}\label{sec3b}} 

We now reflect on the theory presented in Ref. \cite{dubbeldam}.

We have defined $\tau_{d}$ as the pore-blockade time in experiments;
i.e., if we define a state of the polymer with $s(t)=0$ as `0'
(polymer just detached from the pore on one side), and with $s(t)=N$
as `N', then $\tau_{d}$ is the first passage time required to travel
from state 0 to state N \textit{without\/} possible reoccurances of
state 0. In Ref. \cite{dubbeldam}, the authors attach a bead at the
$s=0$ end of the polymer, preventing it from leaving the pore; and
their translocation time ($\tau_{v}$ hereafter) is defined as the
first passage time required to travel from state 0 to state N
\textit{with\/} reoccurances of state 0. This leads them to express
$\tau_{v}$ in terms of the free energy barrier that the polymer
encounters on its way from state 0 to $s=N/2$, where the polymer's
configurational entropy is the lowest. Below we settle the differences
between $\tau_{v}$ of Ref. \cite{dubbeldam} and our $\tau_{d}$.

Consider the case where we attach a bead at $s=0$ and another at
$s=N$, preventing it from leaving the pore. Its dynamics is then
given by the sequence of states, e.g., 
\begin{eqnarray}
...\mbox{N \!x\! m\! x}\overbrace{\mbox{0 \!x$'$0
      \!x$'$m$'$x$'$m$'$x$'$0 \!x$'$}\underbrace{\mbox{0 \!x \!m \!x
	\!m \!x \!m \!x
	\!N}}_{\tau_{d}}}^{\tau_{v}}\mbox{x$'$N}...
\nonumber
\end{eqnarray}
where the corresponding times taken ($\tau_{v}$ and $\tau_{d}$) are
indicated. At state x and x$'$ the polymer can have all values of $s$
except $0,N/2$ and $N$; and at states m and m$'$, $s=N/2$.  The
notational distinction between primed and unprimed states is that a
primed state can occur only between two consecutive states 0, or
between two consecutive states N, while an unprimed state occurs only
between state 0 and state N. A probability argument then leads us to
\begin{eqnarray}
\frac{\tau_{v}}{\tau_{d}}\,=\,\frac{1}{p_{\text{x}}+p_{\text{m}}}\,=\,\frac{f_{\text{x}}\,(1+f_{\text{m}})}{(p_{\text{m}}+p_{\text{m$'$}})f_{\text{m}}(1+f_{\text{x}})}\,,
\label{e1}
\end{eqnarray}
where $p_{\text{m}}$, $p_{\text{m}'}$ and $p_{\text{x}}$ are the
probabilities of the corresponding states,
$f_{\text{m}}=p_{\text{m}}/p_{\text{m}'}$ and
$f_{\text{x}}=p_{\text{m}}/p_{\text{x}}$. Since the partition sum of a
polymer of length $n$ with one end tethered on a membrane is given by
$Z_{n}\sim\lambda^{n}~n^{\gamma_{1}-1}$ with $\lambda$ a non-universal
constant and $\gamma_{1}=0.68$ \cite{diehla}, we have
$p_{\text{m}}+p_{\text{m}'}=Z_{N/2}^{2}/\left[\sum_{s=0}^{N}Z_{s}Z_{N-s}\right]\sim1/N$.
Similarly, $f_{\text{x}}\sim1/N$ \cite{wolt}. Finally,
$f_{\text{m}}\approx1$ \cite{note} yields $\tau_{v}\sim\tau_{d}$.

We have thus shown that the free energy barrier does not play a role
for $\tau_{v}$, implying that the theoretical expression for
$\tau_{v}$ in Ref. \cite{dubbeldam} cannot be correct. The numerical
result $\tau_{v}\sim N^{2.52\pm0.04}$ in Ref. \cite{dubbeldam},
however, confirms our theoretical expression $\tau_{d}\sim N^{2+\nu}$.

\section{Conclusion\label{sec4}}

To conclude, we have shown that for the swollen Rouse chain,
translocation is sub-diffusive up the configurational relaxation time
(Rouse time) of the molecule, after which it has a further Fickian
regime before the longer dwell time is exhausted: the mean square
displacement along the chain is described by $\langle\Delta s^2(t)\rangle
\sim t^{(1+\nu)/(1+2\nu)}$ up to $t\sim N^{1+2\nu}$, after which
$\langle\Delta s^2(t)\rangle \sim t$.  Consequently, the mean dwell time
scales as $\tau_d \sim N^{2+\nu}$. 

In future work, we will
study the role of hydrodynamics. Rouse friction may be an appropriate
model for the dynamics of long biopolymers in the environment within
living cells, if it is sufficiently gel-like to support screened
hydrodynamics on the timescale of their configurational
relaxation. However, we should also ask what is expected in the other
extreme of unscreened (Zimm) hydrodynamics. For our theoretical
discussion the key difference is that, instead of the Rouse time
$\tau_{R}$, in the Zimm case the configurational relaxation times
scale with $N$ according to $\tau_{\text{Zimm}}\sim N^{3\nu}$ in 3D,
which upon substitution into our earlier argument would gives the
lower bound value $\alpha=(1+\nu)/(3\nu)$ for the time exponent of the
impedance, leading to $\tau_{d}\sim N^{1+2\nu}$ (whose resemblance to
the Rouse time is a coincidence --- note that with hydrodynamics Rouse
time loses all relevance). These results, however, do need to be
verified by simulations incorporating hydrodynamics.
%
%
%

%

\begin{thebibliography}{99}
\bibitem{aka} The exponent $\nu$ is also known as the Flory exponent
  and sometimes also referred to as the ``swelling exponent''.
\bibitem{strict} Strictly speaking, $\tau_{R}$ in this expression
should be replaced by the characteristic equilibration time of a tethered
polymer with length of $O(N)$; since both scale as $N^{1+2\nu}$,
we use $\tau_{R}$ here, favouring notational simplicity.
\bibitem{mistake} In Ref. \cite{wolt} we overlooked the
$\tau_{\mbox{\scriptsize M}\looparrowright\mbox{\scriptsize
M}}$ term of Eq. (\ref{eq2}) to (mistakenly) deduce
$\tau_d=2\tau_u$. Nevertheless, due to the
independence of Eq. (\ref{eq4}) on $N$ our results for the scaling of
$\tau_d$ in Ref. \cite{wolt} remain unaffected.
\bibitem{note} $f_{\text{m}}$ is a number slightly smaller than
$1$: if the polymer reaches $s=N/2$ from state 0, due to the memory
effects, it will have slightly higher chance to go back to state 0
rather than to proceed to state N \cite{wolt}.
\end{thebibliography}
\end{document}